\documentclass[12pt, a4paper]{article}
\newtheorem{theorem}{Theorem}

\begin{document}

\title{Quantum Game Theory}
\author{Chiu Fan Lee\thanks{c.lee1@physics.ox.ac.uk} 
\ and \ Neil F. Johnson\thanks{n.johnson@physics.ox.ac.uk}
\\
\\ Center for Quantum Computation and Physics Department \\ Clarendon
Laboratory, Oxford University \\ Parks Road, Oxford OX1 3PU, U.K.}

\maketitle

\abstract{
We pursue a general theory of quantum games. 
We show that quantum games are more efficient 
than classical games, and provide a saturated 
upper bound for this efficiency. 
We demonstrate that the set of finite classical games is a strict subset 
of the set of finite quantum games.
We also
deduce the quantum version of the Minimax Theorem and 
the Nash Equilibrium Theorem.
} \\
\hfill
\\
\newpage

The field of quantum games is currently attracting much attention within the physics
community \cite{Meyer,EWL99,LJ02,qgame}. In addition to their own intrinsic interest, quantum
games offer a new vehicle for exploring the fascinating world of quantum information
\cite{LJ02,qgame, LJ02n}.  
So far, research on quantum games has tended to concentrate on finding
interesting phenomena when a particular  classical game is quantized. As a result, studies of
quantum games have centered on particular special cases rather than on the development of a
general theoretical framework.

This paper aims to pursue the general theory of quantum games. 
We are able to identify a definite sense in which quantum games are `better'
than classical games, in terms of their efficiency. 
We show explicitly that, in terms of the number of 
(qu)bits required, there can be
a factor of 2 increase in efficiency if we play games quantum mechanically.
Hence we are able to quantify a
distinct advantage of quantum games as compared to their classical counterparts. 
Furthermore, through the formalism developed, we are able to 
demonstrate that the set of finite classical games is a strict subset of 
finite quantum games. Namely, all finite classical games can be played by
quantum rules but not vice versa.
We also deduce the quantum version of 
two of the most important theorems in classical game theory:
the minimax theorem for zero-sum games
and the Nash theorem for general static games.

We start by defining what is meant by a game.
In game theory, 
a game consists of a set of players, a set of rules which 
dictate what actions a player can take, and 
a payoff function specifying the reward for a given set of played strategies. 
In other words,
it is a triple $\langle N, \Omega, P \rangle$ where $N$ is the number
of players, $\Omega= \times_k \Omega_k$ with 
$1\leq k\leq N$ such that each set $\Omega_k$ is 
the set of strategies for the $k$-th
player, and $P : \Omega \rightarrow {\bf R}^N$
such that each
$P_k(.)$ with $1\leq k \leq N$ is the payoff function for the $k$-th player. 
Without loss of generality, we can imagine the existence of a referee 
who computes the corresponding payoff function after he
receives the strategies being played by each of the players.
This formal structure includes all classical games and
all quantum games. In other words any game, whether classical or quantum,
is fully described by the corresponding triple
$\langle N, \Omega, P
\rangle$. 
On the other hand, given any triple $\langle N, \Omega, P 
\rangle$, it is not hard to imagine a purely classical game
that might be associated to it.

So in what sense are quantum games any `better' than classical games?
We believe that the answer lies in the issue of efficiency.
Although any game could be played classically, the 
physically feasible ones form a very restricted subset. 
As we have seen, playing a game is ultimately about information exchange
between the players and the referee - hence if a particular game requires you to submit an
infinite amount of information before the payoff functions can be computed, it will not be a
playable game. In other words, we are interested in those games which require only a finite 
amount of resources and time to play.
Hence a connection can be made between this consideration
and the study of algorithms.
In the study of computation, we learn that there are computable functions which
may however not be computed efficiently. Shor's great contribution to information theory was to
advance the boundary of the  set of efficiently
computable functions \cite{Sho97}. This naturally begs the question as to whether it is more
efficient to play games quantum mechanically than classically. We will show shortly
that, in terms of efficiency, some quantum games can indeed outperform classical games.

The quantum game protocol that we study is a generalization of
that described in \cite{EWL99}. We use the term {\it static quantum
games} to reflect the similarity of the resulting games to static classical games. To play a
static quantum game,
we start with an initial state $\rho$ which is represented
by qubits. The referee then
divides the state into $N$ sets of qubit parts, sending 
the $k$-th set to player $k$.
The players separately operate on the qubits that they receive,
and then send them back to the referee.
The referee then determines the payoff for the players with regard to
the measurement outcome of a collection of POVM operators
$\{M_m \}$. Anticipating
the focus on efficiency, we fix the dimension of $\rho$ to be $2^{qN}$
where $qN$ is the number of qubits, and we set $n=2^q$.
We also assume that the players share the initial qubits equally, i.e.
each one of them will receive $q$ qubits. This assumption is inessential
and is for ease of exposition only.
Following the game's protocol, the players operate independently on the states,
and hence it is natural to allow them access to all possible physical maps. Specifically,
we allow each player to have access to 
the set of trace-preserving completely positive maps, 
i.e. for each $k$, we set $\Omega_k$
to be the set of trace-preserving completely positive maps. Indeed,
the only way to restrict the players' strategy sets in this protocol
is to perform some measurement at
the referee's end - this is incorporated into our formalism by allowing the 
referee the set of all POVM operators.

First, we restrict ourselves to two-player games.
If a payoff of $a_m^k$ for player $k$ is
associated with the measurement outcome $m$, the payoff for player $k$
is then tr$(R^k \pi )$ where 
$R^k:=
a_1^k M_1^\dag M_1 + \cdots +
a_{L}^k  M_L^\dag M_L$ and
$\pi$ is the resulting state.
For example, if player I decides to use operation
${\cal E}=\{ E_k\}$ and player II ${\cal F}
=\{ F_k\}$, then $\pi= \sum_{k,l} (E_k
\otimes F_l) \rho (E_k^\dag \otimes F_l^\dag)$. Hence
$P_{\rm I}({\cal E},{\cal F}) = \sum_{k,l} {\rm tr} (R^{\rm I} (E_k
\otimes F_l) \rho (E_k^\dag \otimes F_l^\dag) ) .$
We now fix a set of operators $\{ \tilde{E}_\alpha \}$ which will form a basis
for the set of operators in the state space. If
${\cal E}=
\{ E_k = \sum_\alpha e_{k \alpha} \tilde{E}_\alpha \}$ and
${\cal F}=
\{ F_l = \sum_\alpha f_{l \alpha} \tilde{E}_\alpha \}$, then the payoff is 
\[
\sum_{k,l,\alpha, \beta, \gamma, \delta }
e_{k \alpha} \overline{e_{k \beta}} f_{l \gamma} \overline{f_{l \delta}} A^k_{
\alpha \beta \gamma \delta}
\]
where $A^k_{\alpha \beta \gamma \delta} :=
{\rm tr} [R^k (\tilde{E}_\alpha \otimes \tilde{E}_\gamma)
\rho (\tilde{E}_\beta^\dag \otimes \tilde{E}_\delta^\dag)
]$. 
Letting $\chi_{\alpha \beta} = \sum_k e_{k \alpha} \overline{e_{k \beta}}$ and
$\xi_{\gamma \delta} = \sum_l f_{l \gamma} \overline{f_{l \delta}}$, then
$\chi$ and $\xi$ are positive hermitian matrices with $16^q-
4^q$ independent real parameters.
Note that the number $16^q$ comes from the fact that $16^q$ real parameters are needed
to specify a $4^q \times 4^q$ 
positive hermitian matrix, while $4^q$ comes from the fact that
$\sum_{\alpha, \beta }
\overline{\chi_{\alpha \beta}} \tilde{E}_\alpha^\dag \tilde{E}_\beta = I$
with the assumptions that $\chi$ is positive and hermitian.
This procedure is the same as the so-called {\it chi matrix representation} \cite{NC00}.
For a general matrix $\Xi$, we now observe that $\overline{{\rm tr}(\Xi)}=$ 
tr$(\Xi^\dag)$. Hence we have
$\chi_{\alpha \beta} \xi_{\gamma \delta} A_{
\alpha \beta \gamma \delta} =\overline{
\chi_{\beta \alpha} \xi_{\delta \gamma} A_{
\beta \alpha \delta \gamma}}$. Therefore, the payoff is actually
$\sum_{\alpha, \beta, \gamma, \delta }
{\rm Re}[\chi_{\alpha \beta} \xi_{\gamma \delta} A_{
\alpha \beta \gamma \delta}]$, which is always real as expected. To recap,
the strategy sets for the players are $\{\chi \},
\{ \xi \}$: these are subsets of the set of
positive hermitian matrices such that 
\begin{equation}
\sum_{\alpha, \beta }
\overline{\chi_{\alpha \beta}} \tilde{E}_\alpha^\dag \tilde{E}_\beta = I\ \ , \ \
\sum_{\alpha, \beta }
\overline{\xi_{\alpha \beta}} \tilde{E}_\alpha^\dag \tilde{E}_\beta = I.
\end{equation}
The payoff is given by $\sum_{\alpha, \beta, \gamma, \delta }
{\rm Re}[\chi_{\alpha \beta} \xi_{\gamma \delta} A_{
\alpha \beta \gamma \delta}]$.
As shown above, we may now identify 
$\Omega_k$ to be the set of positive semi-definite hermitian matrices 
satisfying condition (1). It then follows that $\Omega = \times_k \Omega_k$ 
is a convex, compact Euclidean
space.

The above analysis can easily be generalized to $N$-player games.
For a particular $N$-player static game,
$P_k(\vec{\chi})=\sum \chi^1 \cdots \chi^N A^k$ where
$A^k = {\rm tr} [ R^k (\tilde{E} \otimes \cdots \otimes \tilde{E}) \rho
(\tilde{E} \otimes \cdots \otimes \tilde{E})]$
(index summation omitted for clarity).

We can now see a striking similarity between static quantum games and
static classical finite games. The payoff for  
a classical finite two-player game has the form
$\sum_{i,j} x_iA_{ij} y_j$ where $x,y$ belong to some multi-dimensional
simplexes and $A$ is a general matrix---the payoff for a static quantum game is
$\sum_{\alpha, \beta, \gamma, \delta }
\chi_{\alpha \beta} \xi_{\gamma \delta} A_{
\alpha \beta \gamma \delta}$ where $\chi, \xi$ belong to some multi-dimensional
compact and convex sets $\Omega_k$. 
Indeed the multi-linear structure of the payoff function and the 
convexity and compactness of the strategy sets, are the essential features underlying
both classical and quantum games. And one may
exploit these similarities to extend some
classical results into the quantum domain. Two immediate examples are the Nash equilibrium
Theorem and the Minmax Theorem. To show this, we first need to recall some relevant
definitions. For a vector $\vec{v}=(v_i)_{i\in N}$, we set 
$\vec{v}_{-k}$ to be $(v_i)_{i\in N \setminus \{k \}
}$ and we denote $(v_1, \ldots, v_{k-1}, v'_k, v_{k+1}, \ldots, v_N)$ by
$(\vec{v}_{-k}, v'_k)$.
We also define the set of best replies
for player $k$ to be $B_k(\vec{\chi}_{-k}):=\{ 
\chi_k \in \Omega_k : P_k(\vec{\chi}_{-k},\chi_k) \geq
P_k(\vec{\chi}_{-k},\chi_k'), \forall \chi_k' \in \Omega_k \}$.
We note that for each $k$,
$B_k(\vec{\chi}_{-k})$ is convex and closed in $\Omega_k$ and hence compact.
Using the notion of best reply, we can easily define what a Nash
equilibrium is:
an operator profile $\vec{\chi}$ is a Nash equilibrium if
$\chi_k \in B_k(\vec{\chi}_{-k})$ for all $k$.

\begin{theorem}{{\rm (Quantum Nash Equilibrium Theorem)}} \\
For all static quantum games, at least one Nash equilibrium exists.
\end{theorem}

\noindent
{\bf Proof:}
We know that $\Omega$ is a convex compact subset of a Euclidean space. Since
$B=\times_k B_k$ is an upper semi-continuous point-to-set map which takes
each $\vec{\chi} \in \Omega$ to a convex set $B(\vec{\chi}) \subseteq \Omega$,
the theorem follows from Kakutani's fixed point theorem \cite{Ka41}.
\hfill Q.E.D. \\
\hfill

We now restrict ourselves to two-player zero-sum game, i.e.,
$a^{\rm I}_m = -a^{\rm II}_m$ for all $m$. Trivially,
given any $\chi$, player I's payoff is bounded above by 
\[ 
v(\chi) = \min_{\xi \in \Omega_{\rm II}}
\sum_{\alpha, \beta, \gamma, \delta }
\chi_{\alpha \beta} \xi_{\gamma \delta} A_{
\alpha \beta \gamma \delta}.
\]
Similarly, given any $\xi$, player II's payoff is bounded below by
\[ 
v(\xi) = \max_{\chi \in \Omega_{\rm I}}
\sum_{\alpha, \beta, \gamma, \delta }
\chi_{\alpha \beta} \xi_{\gamma \delta} A_{
\alpha \beta \gamma \delta}.
\]
We therefore define 
\begin{eqnarray*}
v_{\rm I} &:=& \max_{\chi \in \Omega_{\rm I}} 
\min_{\xi \in \Omega_{\rm II}} \sum_{\alpha, \beta, \gamma, \delta }
\chi_{\alpha \beta} \xi_{\gamma \delta} A_{
\alpha \beta \gamma \delta}, \\
v_{\rm II} &:=& \min_{\xi \in \Omega_{\rm II}}  
\max_{\chi \in \Omega_{\rm I}}  \sum_{\alpha, \beta, \gamma, \delta }
\chi_{\alpha \beta} \xi_{\gamma \delta} A_{
\alpha \beta \gamma \delta}.
\end{eqnarray*}

\begin{theorem}{{\rm (Quantum Minimax Theorem)}}
\[
v_{\rm I} = v_{\rm II}.
\]
\end{theorem}

\noindent 
{\bf Proof:}
We have seen that for each $k$,
$\Omega_k$ is compact and convex as a Euclidean space.
Also the payoff is linear and continuous 
in $\chi_{\alpha \beta}$ and $\xi_{\alpha \beta}$. Therefore, the theorem
follows from the Minimax theorem in Ref. \cite{Be97}.
\hfill Q.E.D. \\
\hfill

We note that the proofs of
theorem~1 and theorem~2 are completely analogous to the corresponding
classical proofs. This is because the underlying theorems involved -
Katutani's theorem and the Minmax theorem -  
are general enough to 
allow for compact and convex strategy sets without restricting them to
only be simplexes.
We also note that although quantum games can profit from some nice classical results,
problematic issues  in classical game theory also carry over to quantum games. For
example, one would expect multiple Nash equilibira in general quantum games. Classical
game theorists have invented {\it evolutionary game theory} 
\cite{We95}
to deal with this problem 
and its quantum analogy awaits full development \cite{KJB01}.

If quantum games were merely some replicas of classical games, or vice versa,
the subject of quantum games would not be very interesting. Here we show that quantum games are
more than that, by showing that in comparison with classical games,
not only the set of finite quantum games is strictly
larger, quantum games can also be played more efficiently.
Before we do this, however, we need to discuss how to quantify efficiency in both
classical and quantum cases, and how we then compare the resulting efficiencies. 
We have seen that the quantum strategy set
$\Omega_k$ is a conpact and convex Euclidean space.
Since any compact and convex Euclidean space 
lies inside some $m$-dimensional simplex, 
yet at the same time contains a smaller $m$-dimensional
simplex as a subset where $m$ is unique and equals $16^q-4^q$ for $\Omega_k$.
The dimensionality will be the
quantity we use to gauge efficiency, because it is well-defined and reflects the number of
(qu)bits needed. For example, in order to play a two-player quantum game we need to exchange
$4q$ qubits in total. This is because the referee needs to send $q$ qubits to 
the two players and they then need to send them back. 
The strategy set for each player has dimension $16^q-4^q$.
If the same number of bit-transfers is allowed in a classical game, 
then the strategy set for each player will be a simplex of dimension
$4^q-1$. Therefore, does it mean that we have a factor of 2 increase in efficiency
by playing quantum games? It is true in general, but is not immediately obvious:
although the quantum strategy set has a higher dimension, we do not yet know whether many of
the strategies are redundant or not. 

In order to show that general quantum games are indeed more efficient, 
we first perform some concrete calculations. Since the choice of
$\{ \tilde{E}_\alpha \}$ is arbitrary, we take $\{ \tilde{E}_\alpha\} =
\{ n_{[ij]} \}$ where $n_{[ij]} $ denotes a $n \times n$ 
square matrix such that $(ij)$-entry $=1$ and
all other entries are equal to $0$. Denoting $\alpha$ by $ij$ and using condition (1),
we have the following
restrictive conditions on $\chi$: $\sum_i \chi_{ijij}=1$ and
$\sum_i \chi_{ijil}=0$, where the first two sub-indices represent
$\alpha$ while the latter two represent $\beta$. A further calculation
shows that $n_{[ij]} \otimes n_{[kl]} = n^2_{[(i-1)n+k,(j-1)n+l]}$. 
For arbitrary $R$ and $\rho$, we find the following:
\[
A_{\underbrace{ab}_\alpha \underbrace{cd}_\beta
\underbrace{ij}_\gamma \underbrace{kl}_\delta}
=R_{((c-1)n+k),((a-1)n+i)} \times \rho_{((b-1)n+j),((d-1)n+l)}.
\]
We are now ready to prove the following theorem:

\begin{theorem}
$A_{\alpha \beta \gamma \delta}$ is diagonal
with non-zero diagonal entries for some $\rho$ and $R$.
\end{theorem}

\noindent
{\bf Proof:}
We first note that the diagonal elements of $A$ are
\[
A_{abcdabcd}=R_{((n+1)d-n),((n+1)a-n)} \times \rho_{((n+1)b-n),((n+1)c-n)}.
\]
Therefore for all $a,b,c,d$, we set
\begin{eqnarray*} 
R_{((n+1)d-n),((n+1)a-n)} & = & 1/n, \\
\rho_{((n+1)b-n),((n+1)c-n)} & = & 1/n.
\end{eqnarray*}
We then set
all the other entries of $R$ and $\rho$ to be 0.
\hfill Q.E.D. \\
\hfill

\noindent [We note that the above construction still holds true in multi-player games. $A$ will
be a tensor with vanishing entries, except for entries with identical indices.]
Any two operations by player I,
$\chi$ and  $\chi'$, are redundant if 
$P_{\rm I}(\chi, \xi)=
P_{\rm I}(\chi' , \xi)$ for all
$\xi$. However in the above game,  $P_{\rm I}(\chi, \xi)=
\sum_{\alpha, \beta} {\rm Re}[\chi_{\alpha
\beta} \xi_{\alpha \beta}] \neq P_{\rm I}(\chi', \xi)=
\sum_{\alpha, \beta} {\rm Re}[\chi'_{\alpha
\beta} \xi_{\alpha \beta}]$ 
in general.
Therefore, the payoff depends on {\em all} of the
independent parameters and there are $16^q-4^q$ of them.
Hence, the upper bound on efficiency is indeed
saturated. One could also envisage varying $\rho$ and $R$ infinitesimally 
to provide a continuum of quantum games with superior efficiency.

We now provide an example.
We consider a two-qubit two-player zero-sum game, and take 
\[
\rho= 
\left( \begin{array}{cccc}
1/2&0&0&1/2 \\ 0&0&0&0 \\ 0&0&0&0 \\1/2&0&0&1/2
\end{array} \right) ;\ \ 
R =\left( \begin{array}{cccc}
1/2&0&0&1/2 \\ 0&0&0&0 \\ 0&0&0&0 \\1/2&0&0&1/2
\end{array} \right).
\]
This means that $\rho = |\psi\rangle \langle \psi |$ where
$|\psi \rangle = \frac{1}{\sqrt{2}}
(|00\rangle +|11 \rangle)$
and $R= |\psi\rangle \langle \psi |$.
For example, the referee may do a von Neumann measurement on the state 
with respect to 
the orthonormal basis $\{ |\psi \rangle,$ $|01\rangle,$ $|10 \rangle,$
$\frac{1}{\sqrt{2}}
(|00\rangle -|11 \rangle) \}$; and then award a payoff of 1 for each player if the 
outcome is $|\psi\rangle$, and a payoff of 0 for any other outcomes.
The strategy set in this game is of dimension 12, 
which corresponds to13 independent strategies classically---therefore 
at least 8 bits have to be
transferred. This is in contrast with the fact that only 4 qubits needed to be transferred
in the quantum version.
The above game, although reasonably simple, does therefore highlight the
potential of quantum games. 

Besides efficiency, we will now show that quantum games are indeed strictly more general
as claimed.
Firstly of all, we note that the classical strategy set and the quantum strategy set
cannot be made identical if the linearity of the payoff function is to be preserved.
This is because there is no linear homeomorphism that maps $\Omega_k$ to a simplex
of any dimension. In essense the positivity of $\Omega_k$, 
i.e, the conditions $\chi_{\alpha
\alpha} \chi_{\beta \beta} \geq |\chi_{\alpha \beta}|^2$ for all
$\chi \in \Omega_k$, spoils this
possibility. Therefore, if we identify $\Omega_k$ as some multi-dimensional
simplex, we must lose linearity of the payoff function. On the other hand, 
we have seen that every strategy 
in $\Omega_k$ is non-redundant in the above games, it is therefore
impossible to play the above game classically no matter how you 
enlarge the strategy set (still finite, of course).

Conversely, we show by
an example how classical games can be played within the formalism of quantum games.
We consider a general two-player two-move game, the game being classical naturally 
suggests that the initial qubits
are not entangled and the payoffs are determined by measuring the resulting
qubits with respect to the computational basis. Hence, without loss of generality,
\[
\rho= 
\left( \begin{array}{cccc}
1&0&0&0 \\ 0&0&0&0 \\ 0&0&0&0 \\0&0&0&0
\end{array} \right) \ {\rm and }  \ 
R^k =\left( \begin{array}{cccc}
r^k_{11}&0&0&0 \\ 0&r^k_{12}&0&0 \\ 0&0&r^k_{21}&0 \\0&0&0&r^k_{22}
\end{array} \right).
\]
As a result,
\[
A^k_{abcdijkl} =
\left\{ \begin{array}{ll} 
R^k_{2(c-1)+k,2(a-1)+i} & {\rm if } \ b=d=j=l=1, a=c, i=k \\
0 & {\rm otherwise.}
\end{array} \right. 
\]
So the only relevant dimensions for player I are
$\chi_{1111}$ and $\chi_{2121}$.
We now recall the conditions imposed on $\chi$, which are
$\sum_i \chi_{ijij}=1$ and $\sum_i \chi_{ijil} =0$. We therefore see that
$\chi_{1111}+\chi_{2121} =1$. Similarly, $\xi_{1111}+\xi_{2121} =1$.
Therefore, playing the above quantum game is the same as playing the 
classical game with the game
matrix below:
\[
A^k= 
\left( \begin{array}{cc}
r^k_{11}&r^k_{12} \\ r^k_{21}&r^k_{22}
\end{array} \right).
\]

In summary we have shown that
playing games quantum mechanically can be more efficient, and have given a
saturated upper bound on the efficiency. In particular, there is a factor of 2
increase in efficiency. We have also
deduced the quantum version of the minimax theorem for zero-sum games and
the Nash theorem for general static games. In addition, we have shown
that finite classical games consist of a strict subset of finite quantum games.
We have also pointed out
the essential characteristics shared by static quantum  and static classical 
games---these are 
the linearity of the payoff function and the convexity and compactness
of the strategy sets. Indeed, the success of using linear programming to
search for Nash equilibira in classical two-player zero-sum games 
relies on these characteristics---one 
would suspect the same method could be applicable to 
the quantum version as well \cite{LJ02n}.  Our final words would be a cautious
speculation on the possibility that the physical and natural world 
might already be exploiting
this efficiency advantage on the microscopic scale \cite{Tu99}.

\vskip0.5in CFL thanks NSERC (Canada), ORS (U.K.) and
Clarendon Fund (Oxford) for financial support.

\end{document}